\documentclass[prb,twocolumn,showpacs,preprintnumbers,amsmath,amssymb]{revtex4}
\usepackage{graphicx}
\usepackage{dcolumn}
\usepackage{bm}
\begin{document}
\title{\bf Stability and electronic structure of the $(1\times 1)$ SrTiO$_3$(110) polar surfaces by first principles calculations}
\author{Fran\c{c}ois Bottin, Fabio Finocchi and Claudine Noguera}
\affiliation{Groupe de Physique des Solides, Universit\'es Paris 6 -
Paris 7 et UMR CNRS 7588,\\
2 place Jussieu, 75251 Paris cedex 05, France}
\date{\today}
\begin{abstract}
The electronic and atomic structure of several $(1\times 1)$ terminations of the (110) polar orientation of SrTiO$_3$ surface are systematically studied by first-principles calculations.
The electronic structure of the two stoichiometric SrTiO- and O$_2$-terminations are characterized by marked differences with respect to the bulk, as a consequence of the polarity compensation. In the former, the Fermi level is located at the bottom of the conduction band, while in the latter the formation of a peroxo bond between the two surface oxygens results in a small-gap insulating surface with states in the gap of the bulk projected band structure.
We also consider three non stoichiometric terminations with TiO, Sr and O compositions, respectively, in the outermost atomic layer, which automatically allows the surface to be free from any macroscopic polarization. They are all insulating.
The cleavage and surface energies of the five terminations are computed and compared, taking into account the influence of the chemical environment as a function of the relative richness in O and Sr. From our calculations it appears that some (110) faces can even compete with the TiO$_2$ and SrO terminations of the (100) cleavage surface: in particular, the (110)-TiO termination is stable in Sr-poor conditions, the (110)-Sr one in simultaneously O- and Sr-rich environments. The available experimental data are compared to the outcomes of our calculations and discussed.
\end{abstract}
\maketitle


\section{Introduction}

Strontium titanate is a material that has many potential applications. Its high dielectric constant makes it a candidate to replace silicon dioxide in some nano-electronic devices. SrTiO$_3$ has been used in photo-electrolysis and as a substrate for the growth of high $T_c$ superconductors as well as of other thin oxide films~\cite{NogueraOx}.
Like other perovskite compounds, SrTiO$_3$ shows a rich physical behavior. It undergoes an anti-ferro-distortive (AFD) transition at $T<105 K$: the cubic structure turns into a tetragonal phase where the neighboring TiO$_6$ octahedra are tilted with opposite angles around a [100] direction, doubling the unit cell. Moreover, SrTiO$_3$ is at the boundary of a ferroelectric (FE) transition~\cite{Itoh,Ravikumar,Pertsev} but remains paraelectric for all temperatures~\cite{ZhongVanderPRB}.

While the (100) surfaces of SrTiO$_3$ have been extensively studied, both theoretically~\cite{Kunc,VanderbiltSSsrtio3,HeifetsPRB},
and experimentally~\cite{Jiang,Kubo,Castell}, the (110) terminations are much less known. Such a scarcity is likely due to the polar character of the (110) orientation. The sequence of atomic layers of O$_2$ and SrTiO stoichiometry implies a monotonic raise of the microscopic electric field, which has to be compensated either through a modification of the surface composition - which leads to non-stoichiometric terminations - or by an anomalous filling of the surface electronic states - which must imply crucial variations of the electronic structure of the surfaces that should be in principle detectable by experiments.

Nevertheless, \{110\} terminations of strontium titanate has been observed quite often, but the effort towards the precise characterization of their atomic-scale morphology and the detailed study of the corresponding electronic structures started only in the last decade. For instance, it is worth noting the investigations by atomic force microscopy~\cite{Szot}, scanning tunnelling microscopy (STM), Auger spectroscopy and low energy electron diffraction(LEED) measurements~\cite{Zegen}, as well as ultraviolet photoemission and x-ray photoemission spectroscopies (UPS and XPS, respectively), coupled to LEED~\cite{Bando1}. In the latter works the authors made a considerable effort to characterize the modifications of the electronic structure that are connected to different preparation conditions of the surface through STM and scanning tunnelling spectroscopy (STS)~\cite{Bando2}, angle resolved photoemission~\cite{Bando3} and conduction measurements~\cite{Bando4}. Not all these studies provide a unifying picture of the surface, which indeed shows a great sensitivity to the thermodynamic conditions, mainly temperature and oxygen partial pressure.

Motivated by that experimental work, some theoretical studies of SrTiO$_3$(110) recently appeared. The optimized surface geometry was obtained through interatomic forces that were derived either by an empirical shell-model~\cite{HeifetsSS}, or by more accurate treatments that are based on the Hartree-Fock approximation, still describing the hamiltonian matrix on semi-empirical grounds~\cite{ArianaSS,Stashans}. The approximate nature of those approaches allowed some trends to be deduced, but it did not permit in most cases to draw quantitative conclusions that could be compared to experiments. On the other hand, first-principle approaches that are based on the Density Functional Theory (DFT) have been widely applied to the study of oxide surfaces in recent years. More specifically to perovskite surfaces, such an approach has been showed to be reliable and complementary to experiments~\cite{VanderbiltSSsrtio3}. It is the aim of the present paper to provide an application of state-of-the-art first-principle methods to the (110) polar termination of SrTiO$_3$, which is the first one to our knowledge.

In particular, we focus on two issues: \textit{(i)} Which is the mechanism at work to cancel the macroscopic polarization? and \textit{(ii)} Is the thermodynamic stability of the (110) termination comparable to that of the (100) cleavage face? Regarding \textit{(i)}, while in the wide-gap rock-salt compounds MgO and NaCl the non-stoichiometric reconstructions such as the octopolar one are clearly favored over polarity compensation by an anomalous filling of surface states, it has been recently suggested for the less ionic ZnO$(0001)$ and ZnO$(000\bar 1)$ surfaces~\cite{Wander,Carlsson} that the latter mechanism may be at work. The coexistence of the rather ionic Sr-O bonds with the more covalent Ti-O ones, as well as the fairly large dielectric constant of SrTiO$_3$ might situate strontium titanate at the borderline between the two previous cases. Regarding \textit{(ii)}, the quite numerous observations of \{110\} $(1\times 1)$ terminations suggest that these surfaces may be obtained in appropriate thermodynamic conditions. However, the ternary nature of strontium titanate, as well as the lack of precise experimental indications on the surface composition, make the first-principle modelling a rather delicate task. Therefore, in this work, we restrict ourselves to several terminations, stoichiometric or not, of $(1\times 1)$ reconstructions, completing a previous preliminary report~\cite{BotFinNog}: the two stoichiometric O$_2$ and SrTiO faces, and the three non-stoichiometric O, TiO and Sr terminations. Their atomic and electronic structures are computed and discussed in Section III. Moreover, their relative stabilities are compared as a function of the chemical environment (see Sections IV and V).


\section{Computational Method}
\label{Computational}

The calculations are carried out within the density functional theory (DFT)~\cite{HoKo}. The exchange and correlation energy is treated via the local density approximation (LDA) using the Perdew-Wang parametrization~\cite{PW92}. The Kohn-Sham orbitals are expanded on a plane wave (PW) basis set and an energy cutoff of 30 Hartree (Ha) is employed. For all calculations, we use the ABINIT computer code~\cite{ABINIT}.

In conjunction with the use of a PW basis set, we adopt
soft norm-conserving pseudopotentials that are generated following
the Troullier-Martins scheme~\cite{TrouMart} in the Kleinman-Bylander
form~\cite{KleiBil} in order to avoid to take into account the inner atomic electrons in the self-consistent cycle explicitly. 
Along the pseudopotential generation process, atom reference states 3$d^1$4$s^{0.5}$4$p^{0.25}$ for Ti, 4$s^2$4$p^6$5$p^1$ for Sr and 2$s^2$2$p^4$ for O are used. The pseudization radii are
$R_c(s)$=2.45 bohr, $R_c(p)$=2.55 bohr and $R_c(d)$=2.25 bohr for Ti,
$R_c(s)$2.0 bohr, $R_c(p)$=2.5 bohr and $R_c(d)$=2.2 bohr for Sr and
$R_c(s)=R_c(d)$=1.4 bohr and $R_c(p)$=1.75 bohr for the O, with $s$,
$p$ and $d$ being the different channels. The Sr 4$s$ and 4$p$
semi-core electrons are treated as valence electrons in the
self-consistent procedure, (which we refer to in the following as the small-core approximation) whereas the Ti 3$s$ and 3$p$ semi-core states are frozen (large-core approximation). Since the latter states have an appreciable superposition with the $4s$ and $3d$ radial orbitals of Ti, in the exchange and correlation functional we take into account the non-linear core correction~\cite{FuchsSch} between the radial density corresponding to the frozen $3s$ and $3p$ Ti orbitals and the valence electron density.

The previously described computational scheme is tested on the cubic phase of bulk SrTiO$_3$. The lattice parameter $a_0$, the bulk modulus $B_0$ and the cohesive energy $E_{coh}$ are very close to the
experimental data (see Table \ref{Tab:Bulkdata}). The large-core approximation for Ti is also checked against the small-core one, and the former is adopted in all slab calculations. The gap between the valence and conduction band computed through the difference of the Kohn-Sham eigen-energies (2.1 eV) is smaller than the experimental value (3.2 eV)~\cite{Cardona}, which is the common case for the DFT~\cite{Godby}.

\begin{table}[h]
\caption{Experimental and computed lattice parameter $a_0$, bulk modulus $B_0$ and cohesive energy $E_{coh}$ for the cubic SrTiO$_3$ phase. The experimental values of $a_0$ and E$_{coh}$ are quoted in Ref.~\cite{Weyrich} and B$_0$ is taken from Ref.~\cite{Fischer}. Two distinct calculations adopting the 
small-core and large-core approximations for the Ti pseudopotential (see text) are also shown.}
\begin{center}
\begin{tabular}{cccc}
\hline
\hline
& Calc. (large-core) & Calc. (small-core) & Experiment \nonumber \\
\hline
$a_0$ ($\AA$)&3.951(+1.2$\%$)&3.875(-0.8$\%$)&3.903 \nonumber \\
$B_0$ (Gpa)&187(+2$\%$)&192(+5$\%$)&183 \nonumber \\
$E_{coh}$ (eV)&34.06(+7$\%$)&37.00(+17$\%$)&31.7 \nonumber \\
\hline
\hline
\end{tabular}
\end{center}
\label{Tab:Bulkdata}
\end{table}

The calculated formation energies of bulk Ti and Sr from the isolated atomic species are: $E_{\rm Ti}^f=4.58 eV$ and $E_{\rm Sr}^f=1.73 eV$, which differ by -6$\%$ and +2$\%$ from the experimental values, respectively. On the other hand, the O$_2$ formation energy ($E_{\rm O_2}^f=7.58 eV$)
is about 45$\%$ greater than the experimental value, like in previous LDA calculations~\cite{GunJones}. 

The surfaces are described in the framework of the slab model. 
The calculations are carried out by sampling the irreducible Brillouin zone by a (4,4,2) Monkhorst-Pack mesh~\cite{MonkPark}. By using a (5,5,2) mesh, the computed total energies decrease by less than 10 meV for the insulating terminations and 20 meV for the metallic ones.
In those cases, when an effective Fermi-surface smearing is employed to get a better convergence~\cite{Mermin}, a careful procedure is used to extrapolate convergent energies in the limit $T \rightarrow 0$.  The stable surface configurations are obtained by minimizing the Hellmann-Feymann forces in the Born-Oppenheimer
approximation, generally starting from the geometry of the ideal unrelaxed surfaces.

In order to avoid spurious interaction between periodic slabs via 
dipole-dipole interactions~\cite{Neug,Beng}, the compositions of the adopted slabs are symmetric upon inversion along the surface normal. Well converged slab total energies can be routinely obtained for typical vacuum widths ranging between 10 and 12 {\AA}. On the other hand, the determination of the minimal number of atomic layers that is allowed needs some care. For instance, a nine-layers slab is sufficient for the two non-stoichiometric (110)-Sr and (110)-TiO terminations, whereas 11
layers are at least needed for the two (110)-O$_2$ and (110)-SrTiO
stoichiometric terminations. In the case of the O-terminated
surfaces, a very careful choice of the bulk reference energy
is also needed, as detailed in the following. To this purpose, we made calculations up to 15-layers thick slabs. 
As a general rule, the slab is considered to be thick enough when
the difference between the total energies of two slabs differing by a SrTiO$_3$ unit and the bulk total-energy is smaller than 1 mHa \endnote{ The uncertainty associated to this value, because of the
number of bulk layers considered and the convergence with
respect to the number of k points, is smaller than the
thickness convergence one.}.
This condition cancels out the problem of nonconvergent
surface energies~\cite{Boettger,Meth} and makes it possible to obtain a fair bulk reference energy for each termination.

Even if our calculations are carried out at 0 K, we impose symmetry constraints to our slabs to simulate the cubic phase that is stable at room temperature.  
A special case is represented by the non-stoichiometric (110)-O termination, since the mirror symmetry along the $[\bar 1 1 0]$ direction is lost. As it will be detailed in the section devoted to such a termination, the bulk energy should be computed with reference to an anti-ferro-distortive (AFD) phase that is simulated in a tetragonal supercell analogous to that used for the slab.
Other authors gave details about the change of the AFD distortion with respect to the bulk structure~\cite{SaiVander}.
Our computed total-energy differences between the cubic and the AFD phases
is about 2 mHa. Neglecting such a tiny quantity may appreciably bias the numerical extrapolation of the surface energy when a linear scheme is used. To summarize, the uncertainty on the surface energies $E_{cl}$ within our slab approach does not exceed $10^{-2}$ J/m$^2$, which is usually much smaller than the computed surface energy differences for the various terminations.

We also focus on the total electron distribution in the various surface configurations. 
An estimate of the electron sharing between the O anions and the Sr and Ti cations is obtained by means of the Bader's topological analysis of the electron density, which corresponds to a partition of the total charge in atomic bassins. Such a procedure is independent of the basis set used~\cite{Bader}, and has been recently used~\cite{Petr} and improved within the ABINIT package. The precision of the atomic charge integration is about $5 \times 10^{-3}$ electrons.


\section{Electronic and atomic structure}

In the (110) orientation of SrTiO$_3$, a stacking sequence
of atomic layers ...O$_2$-SrTiO-O$_2$-SrTiO...  is provided.
If we consider that the ionic charge of O, Ti and Sr are $Q_{\rm O}=-2$,
$Q_{\rm Ti}=+4$ and $Q_{\rm Sr}=+2$ respectively, then the O$_2$ plane bears a formal charge $Q_{\rm O_2}=-4$ and the SrTiO plane $Q_{\rm SrTiO}=+4$ per 2D unit cell. According to the criterion for polarity compensation~\cite{Noguerajpcm,Petr}, the formal surface charges of the various SrTiO$_3$(110) faces have to be equal to half the bulk value, i.e., $Q_{surf}=\pm2$.
Therefore, we consider two main classes of $(1\times 1)$ terminations: on one side, the so called stoichiometric terminations, since their compositions reflects the bulk stacking: the (110)-SrTiO and -O$_2$ ones, for which an anomalous filling of surface states is expected. On the other side, we study the (110)-TiO, -Sr and -O terminations, for which the stoichiometry changes can in principle provide the polarity compensation. For each termination, we describe in the following the detailed atomic and electronic structure. 

\subsection{Stoichiometric terminations}

\subsubsection{The SrTiO termination.}

\begin{figure}[h]
\centering
\mbox{\includegraphics[height=5cm,width=6cm,angle=-90]{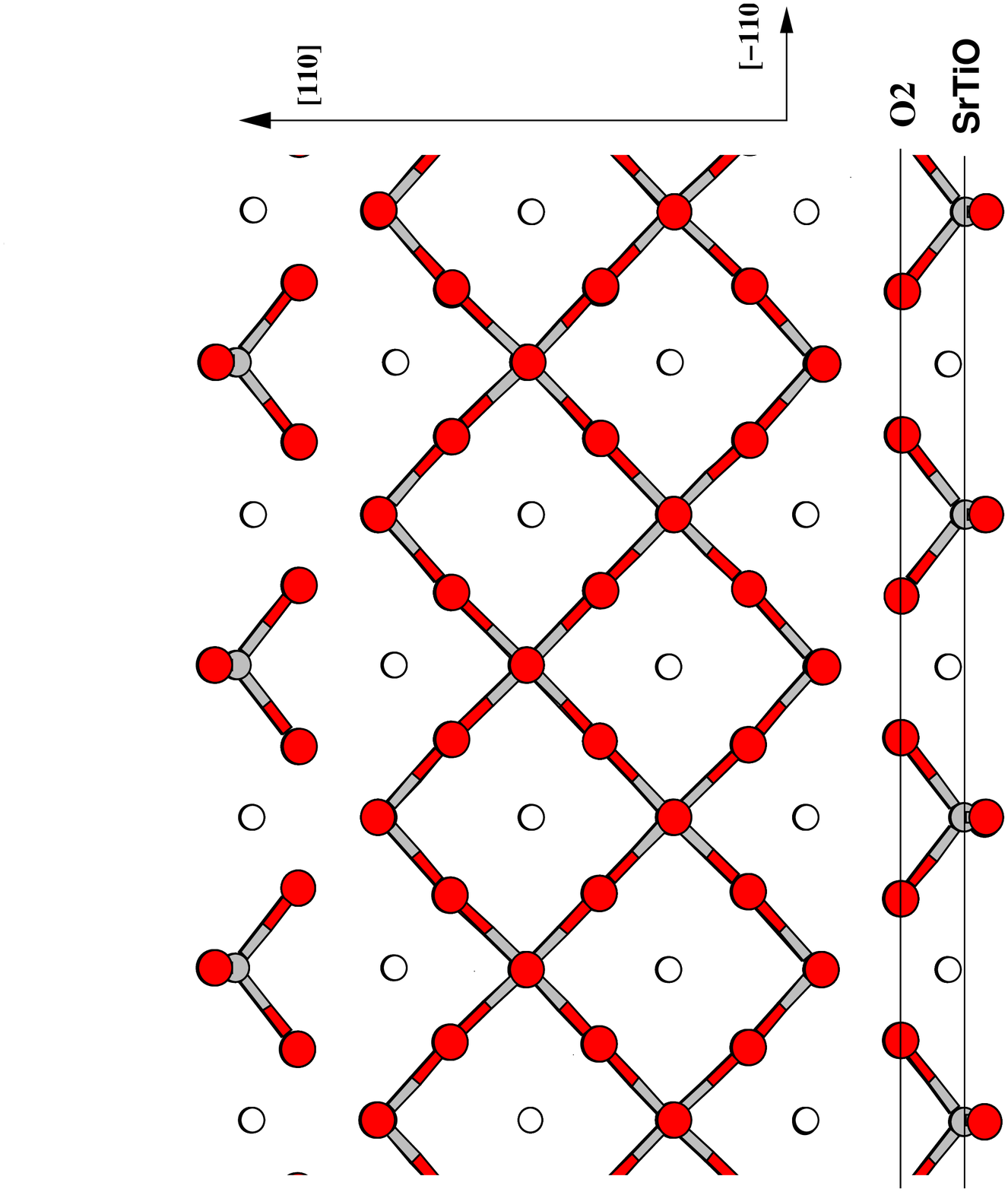}}
\caption{{Side view of the SrTiO slab cut along a [001] plane. Sr, Ti and O
atoms are white, grey and red, respectively.}}
\label{Fig:srtioslab}
\end{figure}

\begin{figure}[h]
\centering
\mbox{\includegraphics[height=4.5cm,width=6cm]{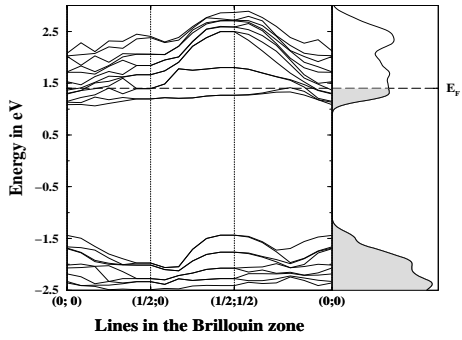}}
\caption{{Computed band structure
for the SrTiO termination of SrTiO$_3$(110). Only the
valence band top and the conduction band bottom are drawn.}}
\label{Fig:band_srtio}
\end{figure}

In the stoichiometric (110)-SrTiO termination (see Figure~\ref{Fig:srtioslab}), 6 Sr-O bonds and 2 Ti-O bonds are cut. 
Since the surface plane bears a formal charge $Q_{\rm SrTiO}=+4$
there must be an anomalous filling of surface states, in order to cancel out the macroscopic component of the slab dipole~\cite{Noguerajpcm}. 
Figure~\ref{Fig:band_srtio} reports the computed band structure. One can note that the Fermi energy is located above the bottom of the conduction band.
Some conduction-like states are thus filled, which bring
the two additional electrons that are needed according to the
electron counting rule for polarity compensation.

\begin{figure}[h]
\centering
\mbox{\includegraphics[height=4.5cm,width=6cm]{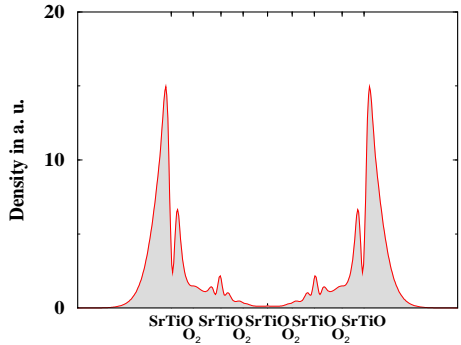}}
\caption{ Localisation of the compensatory state at the
M (0.5;0.5;0.0) point of the Brillouin zone}
\label{Fig:loc_M_srtio}
\end{figure}

These states are localized in the outermost layers and are genuine surface states. The density of the filled conduction band state at the M point of the Brillouin zone that is displayed in figure~\ref{Fig:loc_M_srtio}, is roughly delocalized over three surface layers, which shows that the self-consistent charge redistribution is more complex than the model that is based upon the values of the formal ionic charges. A more thorough insight can be obtained by means of the Bader's topological charge analysis, which is reported in Table~\ref{Tab:Bader_srtio}.  
The layer charge q$_{\rm SrTiO}^{I,III}$ and q$_{\rm O_2}^{II}$ of the first and third SrTiO, and the second O$_2$ outermost planes, respectively, are indeed strongly modified by the filling of the additional surface state and thus show marked changes from the computed charge for the inner bulk-like layers (q$^{Bulk}=\pm 2.50$), which is roughly achieved only in the fourth layer. 
The polarity compensation criterion is fulfilled, since the sum of the charge of the four first layers is equal to -1.30, which is almost the half of the bulk layer charge q$^{Bulk}$.

\begin{table}[h]
\caption{ q$_{\rm O}$, q$_{\rm Ti}$ and q$_{\rm Sr}$ are the Bader's
topological charge of the Oxygen, Titanium and Strontium respectively,
for the SrTiO termination.
I, II, III and IV give the index of the layer. Charge are given in
electrons. For comparisons, bulk values are given at the bottom.}
\begin{center}
\begin{tabular}{c|c|c}
\hline
\hline
layer & Atomic charges & Layer charge \nonumber \\
\hline
SrTiO & q$_{\rm O}^{I}$=-1.36; q$_{\rm Ti}^{I}$=1.74; q$_{\rm Sr}^{I}$=1.40 &
q$_{\rm SrTiO}^{I}$=1.78  \nonumber \\
O$_2$ & q$_{\rm O}^{II}$=-1.38 & q$_{\rm O_2}^{II}$=-2.76  \nonumber \\
SrTiO & q$_{\rm O}^{III}$=-1.30; q$_{\rm Ti}^{III}$=2.03; q$_{\rm Sr}^{III}$=1.55 &
q$_{\rm SrTiO}^{III}$=2.28  \nonumber \\
O$_2$ & q$_{\rm O}^{IV}$=-1.30 & q$_{\rm O_2}^{IV}$=-2.60  \nonumber \\
\hline
Bulk & q$_{\rm O}^{Bulk}$=-1.26; q$_{\rm Ti}^{Bulk}$=2.18; q$_{\rm Sr}^{Bulk}$=1.58 &
q$^{Bulk}$= 2.50 \nonumber \\
\hline
\hline
\end{tabular}
\end{center}
\label{Tab:Bader_srtio}
\end{table}

\begin{figure}[h]
\centering
\mbox{\includegraphics[height=9cm,width=9cm,angle=-90]{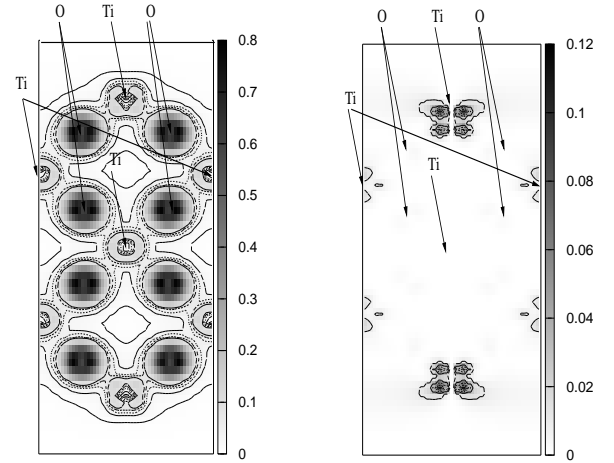}}
\caption{Cutting of the total valence density (left) and compensatory 
state density (right) perpendicularly to the [100] direction, in a TiO$_2$ plane}  
\label{Fig:Cut_srtio_tio2}
\end{figure}

Finally, we can see in table~\ref{Tab:Bader_srtio} that the
difference between the surface layer charge q$_{\rm SrTiO}^{I}$ and
the bulk layer charge q$^{Bulk}$ is mainly localized
on the Titanium, which is confirmed by the direct visualization of the filled state at the conduction band bottom (in the right part of figure~\ref{Fig:Cut_srtio_tio2}).
However, an effective decrease of the electronic kinetic energy may be obtained through the delocalization of this state over neighboring sites. Such a screening phenomenon, reminiscent to that found in Na-covered TiO$_2$(110) surfaces~\cite{Tristanth}, thus affects the Sr and O charges, too.
The charge of the outermost Ti is thus reduced, which can be probed by XPS experiments through a surface Ti signal that should differ from the bulk Ti$^{4+}$ formal oxidation state to a great extent. 

\begin{table}[h]
\caption{ Relaxation and rumpling on the (110)-SrTiO termination. The mean positions of the SrTiO layers are computed by averaging the normal coordinates of the corresponding atoms. The interplanar distances are
given in {\AA} and their relative variations in brackets.}
\begin{center}
\begin{tabular}{c|c}
\hline
\hline
layer & Relaxations and rumplings  \nonumber \\
\hline
SrTiO & Ti(-0.03) O(+0.38) Sr(-0.35) \nonumber \\
$\updownarrow$ & 1.25 (-11$\%$)  \nonumber \\
O$_2$ & no rumpling \nonumber \\
$\updownarrow$ & 1.66 (+19$\%$)  \nonumber \\
SrTiO & Ti(+0.06) O(+0.13) Sr(-0.19)\nonumber \\
$\updownarrow$ & 1.28 (-8$\%$)  \nonumber \\
O$_2$ & no rumpling \nonumber \\
$\updownarrow$ & 1.49 (+6$\%$)  \nonumber \\
SrTiO & Ti(+0.01) O(+0.04) Sr(-0.05) \nonumber \\
$\updownarrow$ & 1.38 (-1$\% $)  \nonumber \\
O$_2$ & no rumpling \nonumber \\
\hline
\hline
\end{tabular}
\end{center}
\label{Tab:Relax_srtio}
\end{table}

As far as the atomic structure is concerned, we can see in 
table~\ref{Tab:Relax_srtio} that large relaxations happen
in the four outermost layers. For instance, the large rumpling between 
Oxygen and Strontium that is equal to 0.73 {\AA} in the
surface layer, is still non negligible in the third layer (0.32 {\AA}).
The inter-planar distance between the outermost 
SrTiO layer and the O$_2$ layer beneath contracts by 11 \%, while that between
the latter and the third SrTiO planes grows by +19\%, showing a typical
damped oscillation that makes the surface unlike layers closer two-by-two,
which is accounted for by the general theory~\cite{DejSpan2,Allan}. However, the surface Ti-O bondlengths $d_{\rm Ti-O}$ are only weakly modified: for Ti in the topmost layer, $d_{\rm Ti-O}$ is reduced of -3\%, while it is expanded by 5\% in the third layer. Such a behavior is due to the combined relaxation and rumpling that mainly rotate the Ti-O bond. At odds, the more ionic and weaker Sr-O bonds are much more affected: $d_{\rm Sr^I-O^{III}}$ is reduced by 13\%, while 
$d_{\rm O^I-Sr^{III}}$ is expanded by 24\%

\subsubsection{The O$_2$ termination.}

\begin{figure}[h]
\centering
\mbox{\includegraphics[height=10cm,width=6cm,angle=-90]{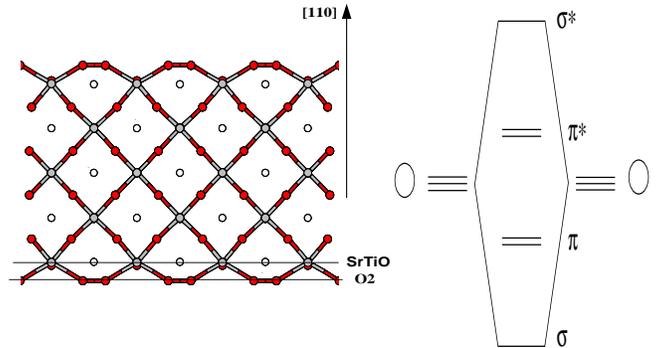}}
\caption{{Left part: side view of the O$_2$-SrTiO$_3$(110)
slab, cut perpendicular to the [001] crystallographic direction.
Right part: schematic molecular diagram of an oxygen dimer. In the case of an isolated O$_2^{--}$ group, the two $\pi^*$ orbitals are completely filled.}}
\label{Fig:peroxo+O2slab}
\end{figure}

The complementary stoichiometric (110)-O$_2$ termination is also expected to undergo deep modifications of the electronic structure because of the polarity compensation, which requires two electrons to be removed from the surface.
As displayed in the left part of Figure~\ref{Fig:peroxo+O2slab},
the two surface oxygens move close to each other forming a bond 1.48 {\AA} long. Such a bond-length is usually a signature of a peroxo group ~\cite{REFO2}, which is formally denoted as ${\rm O}_2^{2-}$ in the ionic limit.
It would imply that the surface charge per unit cell is $Q_{surf}=-2$
instead of -4 like in the bulk. Therefore, the formation of a peroxo group formally permits to fulfill the polarity compensation criterion.


\begin{figure}[h]
\centering
\mbox{\includegraphics[height=5cm,width=6cm]{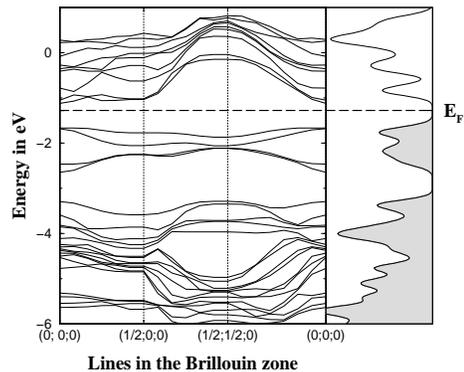}}
\caption{{Computed band structure
for the O$_2$-terminated slab. Only the
valence band top and the conduction band bottom are drawn. The $\pi^*$ states discussed in the text give rise to four bands (two for each termination) which are not completely degenerated due to finite-slab effects.}}
\label{Fig:band_o2}
\end{figure}

\begin{figure}[h]
\centering
\mbox{\includegraphics[height=6cm,width=4cm,angle=-90]{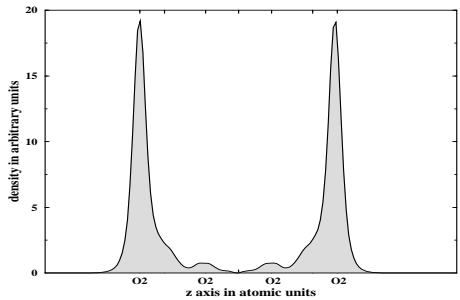}}
\caption{{Localisation of the highest occupied state $\pi^*$
for the O$_2$ termination of SrTiO$_3$(110).}}
\label{Fig:loc_pi_o2}
\end{figure}

Less formally, one can see that the formation of a peroxo bond is made possible by emptying the anti-bonding $\sigma^*$ molecular orbital (see the right panel of Figure~\ref{Fig:peroxo+O2slab}).
This also provides an effective mechanism to open a small gap at the surface (see figure~\ref{Fig:band_o2}). 
The Fermi level rests above the two anti-bonding states $\pi^*$
and does not cross the band structure .
The bulk gap value of SrTiO$_3$ (around 2 eV) is found between
the valence band and the conduction band (i.e. excluding the O$_2$ $\pi^*$ states).
The two in-gap $\pi^*$ states are split since the two $[\bar 110]$ and [001] crystallographic directions are not equivalent, and are localized at the surface layer as displayed in figure~\ref{Fig:loc_pi_o2}.

\begin{table}[h]
\caption{ Same caption as Table~\ref{Tab:Bader_srtio} for O$_2$ termination}
\begin{center}
\begin{tabular}{c|c|c}
\hline
\hline
layer & Atomic charges & Layer charge \nonumber \\
\hline
O$_2$ & q$_{\rm O}^{I}$=-0.66 & q$_{\rm O_2}^{I}$=-1.32  \nonumber \\
SrTiO & q$_{\rm O}^{II}$=-1.20; q$_{\rm Ti}^{II}$=2.13; q$_{\rm Sr}^{II}$=1.55 &
q$_{\rm SrTiO}^{II}$=2.48  \nonumber \\
O$_2$ & q$_{\rm O}^{III}$=-1.21 & q$_{\rm O_2}^{III}$=-2.42  \nonumber \\
SrTiO & q$_{\rm O}^{IV}$=-1.24; q$_{\rm Ti}^{IV}$=2.19; q$_{\rm Sr}^{IV}$=1.58 &
q$_{\rm SrTiO}^{IV}$=2.53  \nonumber \\
\hline
Bulk & q$_{\rm O}^{Bulk}$=-1.26; q$_{\rm Ti}^{Bulk}$=2.18; q$_{\rm Sr}^{Bulk}$=1.58 &
q$^{Bulk}$=-2.50 \nonumber \\
\hline
\hline
\end{tabular}
\end{center}
\label{Tab:Bader_o2}
\end{table}

Table \ref{Tab:Bader_o2} yields the results of a Bader's topological analysis for this termination. The topological charges of the surface O (${\rm O}^I$) are sensitively reduced, consistently with the formation of the peroxo bond. 
We obtain a surface charge almost equal to half the bulk one.
In this case, the charge modification is essentially restricted to the outermost O atoms, as a consequence of the covalent and localized nature of the peroxo bond.

\begin{table}[h]
\caption{ Same caption as Table~\ref{Tab:Relax_srtio} for O$_2$ termination}
\begin{center}
\begin{tabular}{c|c}
\hline
\hline
Layer&Relaxations and rumplings  \nonumber \\
\hline
O$_2$&no rumpling \nonumber \\
$\updownarrow$&1.12 (-20$\% $)  \nonumber \\
SrTiO&Ti(-0.05) O(+0.12) Sr(-0.07) \nonumber \\
$\updownarrow$&1.52 (+9$\% $)  \nonumber \\
O$_2$&no rumpling \nonumber \\
$\updownarrow$&1.32 (-6$\% $)  \nonumber \\
SrTiO& Ti(-0.02) O(+0.03) Sr(-0.01)\nonumber \\
$\updownarrow$&1.45 (-4$\% $)  \nonumber \\
O$_2$&no rumpling \nonumber \\
$\updownarrow$&1.40 (0$\% $)  \nonumber \\
SrTiO&no rumpling \nonumber \\
\hline
\hline
\end{tabular}
\end{center}
\label{Tab:Relax_o2}
\end{table}

As far as the atomic structure is concerned, important modifications
are present on the surface. The Oxygens tilt at the surface
and reduce their angle with respect to the [110] direction by
strongly increasing the bond-length with the Titanium on second
layer ($d_{\rm O^I-Ti^{II}}$) by 19 \%. As a consequence,there is a big inward relaxation (-20$\%$ -- see also table~\ref{Tab:Relax_o2}) of the
peroxo bond with respect to the second layer.
On the other hand, $d_{\rm O^I-Sr^{II}}$ is remarkably reduced (-13 \%).

We point out that a few surface configurations were obtained
for the (110)-O$_2$ termination through the minimization procedure. 
The one reported in Fig.\ref{Fig:peroxo+O2slab} has the lowest total energy.
Another configuration with the peroxo group on top of the
Titanium is found, which is very slightly higher in energy (5.$10^{-3} J/m^2$)
but still within the intrinsic precision of our calculations. 
These two configurations have very similar band structures and atomic charges,
and can be considered as almost degenerated. It is worth noting that when starting the geometry optimization from the ideal, unrelaxed (110)-O$_2$ termination, a third local minimum without peroxo groups at the surface is found. With respect to the peroxo configurations, it has an
open-shell structure and a higher surface energy of about 0.7 J/m$^2$.

\subsection{Non-stoichiometric terminations}

\begin{figure}[h]
\centering
\mbox{\includegraphics[height=6cm,width=9cm]{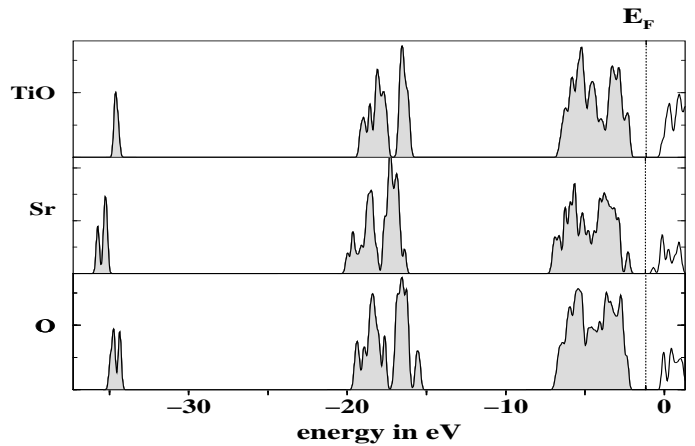}}
\caption{Density of states of the TiO, Sr and O terminations. The
occupied states are filled in gray and $E_F$ is the Fermi level.}
\label{Fig:dos_o_sr_tio}
\end{figure}

The (110)-TiO, -O and -Sr terminations may be ideally obtained from the stoichiometric surface by adsorption or desorption of Sr and O atoms. 
They bear a formal charge $Q_{\rm TiO}=+2$, $Q_{\rm O}=-2$ and $Q_{\rm Sr}=+2$ in the ionic limit, respectively, which corresponds to half the bulk layer charge $Q_{bulk}=\pm 4$. 
Therefore, they are all compensated and no anomalous filling of surface states is in principle necessary.
In order to demonstrate this assumption, the computed density of states of those three non-stoichiometric terminations are drawn in Fig.\ref{Fig:dos_o_sr_tio}.
They are all insulating, with electronic structures qualitatively similar to the bulk.

\subsubsection{The Sr and TiO terminations.}
\begin{figure}[h]
\centering
\mbox{\includegraphics[height=9cm,width=5cm,angle=-90]{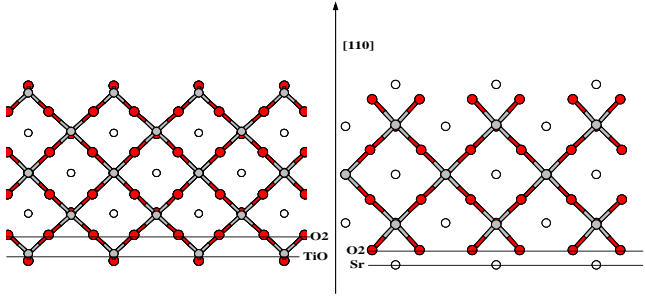}}
\caption{Left (right) panel: side view of the TiO (Sr) slab, cut along a [001] plane.}
\label{Fig:sr+tioslab}
\end{figure}

The (110)-Sr termination is obtained from the O$_2$ one by adsorbing a Sr atom per unit cell whereas the (110)-TiO one is obtained from the (110)-SrTiO termination by removing a row of surface Sr along the [100] direction. These two terminations are displayed in figure~\ref{Fig:sr+tioslab}.

These two (110) terminations are representative of more open surfaces. The coordination numbers of the surface atoms are reduced with respect to the stoichiometric (110) $(1\times 1)$ terminations as well as to the non-polar (100) surfaces.
Let us imagine to cleave a SrTiO$_3$ slab along different orientations: two complementary terminations are obtained, for which a certain number of bonds are missing. Such cut bonds are: 4 Sr-O and 1 Ti-O for the 
non-polar (100) orientation; 2 Ti-O and 6 Sr-O bonds for the (110)-SrTiO
and (110)-O$_2$ terminations, as well as for the (110)-O termination, which is self-complementary; 2 Ti-O and 8 Sr-O bonds for the (110)-TiO and (110)-Sr terminations. Moreover, at variance with the TiO$_2$ and SrO (100) terminations and SrTiO, O$_2$ and O (110) ones, the (110)-Sr and the (110)-TiO terminations are the only ones having an under coordinated atom even in the third layer starting from the surface (a Sr-O bond is cut). Therefore, it is not surprising that the remaining atoms on the (110)-Sr and (110)-TiO undergo large relaxations (see table~\ref{Tab:sr_tio_Relax}), like the inward relaxation of the surface Sr on the Sr termination (-43$\%$) and the strong rumpling of the Ti and O atoms on the TiO termination (0.48$\AA$). 
Despite this under-coordination, relaxations are more quickly damped than for the stoichiometric terminations.
For instance, the inter-planar relaxation between the third and the fourth layer is reduced by a factor two with respect to the stoichiometric (110)-SrTiO and (110)-O$_2$ terminations.

\begin{table}[h]
\caption{ Same as Table~\ref{Tab:Relax_srtio}, for the (110)-TiO and (110)-Sr terminations.}
\footnotesize
\begin{center}
\begin{tabular}{c|c|c}
\hline
\hline
Layer&Relaxations (TiO)&Relaxations (Sr)  \nonumber \\
\hline
TiO/Sr&Ti(-0.24) O(+0.24)&no rumpling   \nonumber \\
$\updownarrow$&1.51 (+8$\%$)&0.79 (-43$\%$)  \nonumber \\
O$_2$&no rumpling&no rumpling  \nonumber \\
$\updownarrow$&1.43 (+2$\%$)&1.53 (+10$\%$)  \nonumber \\
SrTiO&Ti(+.08) O(-.07) Sr(-.01)&Ti(+.06) O(-.02) Sr(-.04)  \nonumber \\
$\updownarrow$&1.32 (-5$\% $)&1.35 (-4$\%$)  \nonumber \\
O$_2$&no rumpling&no rumpling \nonumber \\
$\updownarrow$&1.40 (0$\% $)&1.41 (+1$\%$) \nonumber \\
SrTiO&no rumpling&no rumpling \nonumber \\
\hline
\hline
\end{tabular}
\end{center}
\label{Tab:sr_tio_Relax}
\end{table}

\begin{table}[h]
\caption{Same caption as Table~\ref{Tab:Bader_srtio} for the (110)-TiO termination. For the (110)-Sr termination, the topological charges of the surface atoms are very similar to the bulk ones, from which they differ by 7\% at most.}
\begin{center}
\begin{tabular}{c|c|c}
\hline
\hline
layer & Atomic charges & Layer charge \nonumber \\
\hline
TiO & q$_{\rm O}^{I}$=-1.20; q$_{\rm Ti}^{I}$=2.04 &
q$_{\rm TiO}^{I}$=0.84  \nonumber \\
O$_2$ & q$_{\rm O}^{II}$=-1.11 & q$_{\rm O_2}^{II}$=-2.22  \nonumber \\
SrTiO & q$_{\rm O}^{III}$=-1.20; q$_{\rm Ti}^{III}$=2.21; q$_{\rm Sr}^{III}$=1.58 &
q$_{\rm SrTiO}^{III}$=2.59  \nonumber \\
O$_2$ & q$_{\rm O}^{IV}$=-1.25 & q$_{\rm O_2}^{IV}$=-2.50  \nonumber \\
\hline
Bulk & q$_{\rm O}^{Bulk}$=-1.26; q$_{\rm Ti}^{Bulk}$=2.18; q$_{\rm Sr}^{Bulk}$=1.58 &
q$^{Bulk}$=-2.50 \nonumber \\
\hline
\hline
\end{tabular}
\end{center}
\label{Tab:Bader_tio}
\end{table}

Since the (110)-Sr and the (110)-TiO terminations are intrinsically
compensated by stoichiometry, we may guess that the atomic surface charges
 differ little from the bulk, and that such an effect is mainly
be due to the reduction of the surface atom coordination numbers. In fact,
the Bader's topological analysis (see Table~\ref{Tab:Bader_tio}) confirms the polarity compensation of the two terminations: the charge modification affects essentially the three outermost layers and their sum on these layers is equal to +1.21, which is almost half of the bulk value. Even if these changes
are less intense than on the stoichiometric terminations, they extend rather deeply into the slabs, which confirms the correlation with the presence of under-coordinated atoms and the interplay between the atomic relaxations and the electron redistribution.

\subsubsection{The O termination.}

\begin{figure}[h]
\centering
\mbox{\includegraphics[height=7cm,width=5cm,angle=-90]{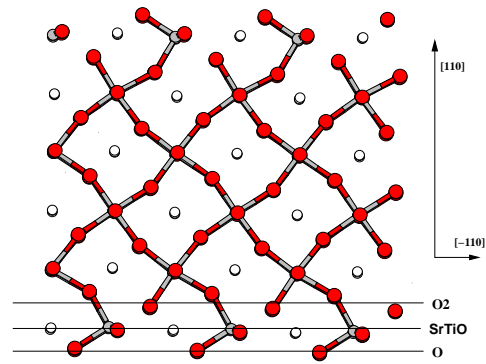}}
\caption{Side view of the (110)-O slab.}
\label{Fig:oslab}
\end{figure}

Such a termination may be ideally obtained in two different ways:
by removing an Oxygen atom from a O$_2$ termination
or by adding an Oxygen atom on a SrTiO termination.
The presence of a single O atom on the surface breaks the mirror symmetry along the $[\bar110]$ direction, at odds with all $(1\times 1)$ terminations that were previously considered. 
Consequently, among all the (110) terminations, the convergence of the atomic structure and the surface energy of this one is the most delicate.
We can see in figure~\ref{Fig:oslab} that the broken surface symmetry permits a distortion and its propagation into the slab. It consists in an alternating rotation of the octahedra along a [100] direction, which is known as an anti-ferro-distortive (AFD) instability. 
We point out that such a abnormal deep propagation of the AFD instability from the surface into the bulk is related to the fact that the low-symmetry phase is stable at T=0 K, the temperature for which the calculations are performed.
This propagation may be hindered by considering a 15 layer slab with the five inner layers frozen. 
Even in that case, the surface atomic structure is practically undistinguishable from that obtained on top of an AFD bulk.

\begin{table}[h]
\caption{ Relaxation and rumpling on the (110)-O termination. The mean positions of the SrTiO layers are computed by averaging the normal coordinates of the corresponding atoms. The inter-planar distances are
given in $\AA$ and their relative variations in brackets.
$\delta_{\rm TiO_2}$ corresponds to the Ti-O lateral relaxation
(the projected interatomic distance along the $[\bar 110]$ direction).}
\begin{center}
\begin{tabular}{c|c|c}
\hline
\hline
Layer&Relaxations and rumpling&$\delta_{\rm TiO_2}$   \nonumber \\
\hline
O&-&- \nonumber \\
$\updownarrow$&0.94 (-33$\% $)&-  \nonumber \\
SrTiO&Ti(-0.09) O(+0.06) Sr(+0.03)&0.391 \nonumber \\
$\updownarrow$&1.42 (+2$\% $)&-  \nonumber \\
O$_2$&O($\pm$0.33)&- \nonumber \\
$\updownarrow$&1.38 (-1$\% $)-  \nonumber \\
SrTiO&Ti(+0.05) O(+0.02) Sr(-0.08)& 0.087 \nonumber \\
$\updownarrow$&1.38 (-1$\% $)&-  \nonumber \\
O$_2$&O($\pm$0.23)&- \nonumber \\
$\updownarrow$&1.376 (-1$\% $)&-  \nonumber \\
SrTiO&Ti(-0.03) O(0.00) Sr(-0.03)& 0.057 \nonumber \\
$\updownarrow$&1.40 (0$\% $)&-  \nonumber \\
O$_2$&O($\pm$ 0.25)&- \nonumber \\
\hline
\hline
\end{tabular}
\end{center}
\label{Tab:o_Relax}
\end{table}

The surface O relaxes inwards (-33$\%$) and laterally toward the two Strontium underneath (see table~\ref{Tab:o_Relax}).
In the inner part of the slab, no large inter-planar relaxation happens.
However, a quite strong rumpling affects the inner O$_2$ layers, which corresponds to the AFD distortion and is not damped as a function of the slab thickness. Moreover, there are large lateral atomic displacements within the SrTiO layers that are denoted with $\delta_{\rm TiO_2}$ (see table~\ref{Tab:o_Relax}) and decrease quite rapidly going into the bulk.
As a consequence of this distortion, the O-Ti-O alignement
is broken along the [001] direction.\\
The topological charges of the surface oxygen's are very similar to the bulk ones, from which they differ by 10\% at most.

\section{Thermodynamic stability}

Because of the different chemical nature of the various  $(1\times 1)$ (110) terminations of SrTiO$_3$, their stability must be discussed as a function of the actual surface composition~\cite{WangPRLal2o3,WangPRLfe2o3,VanderbiltSSsrtio3,ArianaSS}. 
To this purpose, we make use of two distinct physical quantities: the first one corresponds to the energy that is necessary to split the crystal in two parts and create two complementary terminations, which we refer to as the cleavage energy $E_{cl}$. 
The second one is the surface grand-potential $\Omega_s^i$, which is a measure of the excess energy of a semi-infinite crystal exposing a termination with a given composition $i$ in contact with matter reservoirs.

\subsection{Cleavage energy}

When a stoichiometric SrTiO$_3$ slab is ideally cut and the two parts are put apart, two complementary surface terminations are created.
In our case, they are the (110)-SrTiO and the (110)-O$_2$ terminations on one hand, the (110)-TiO and the (110)-Sr ones on the other. 
The (110)-O termination can be considered as self-complementary. 
The respective cleavage energies $E_{cl}^{\rm (O_2 + SrTiO)}$ and $E_{cl}^{\rm (TiO + Sr)}$ can be obtained from the total energies computed for the symmetric slabs through the following equations:
\begin{eqnarray}
E_{cl}^{\rm (TiO + Sr)} & = & \frac{1}{2S}( E_{slab}^{\rm TiO} + E_{slab}^{\rm Sr} - nE_{bulk} ) \\
E_{cl}^{\rm (O_2 + SrTiO)} & = & \frac{1}{2S}( E_{slab}^{\rm O2} + E_{slab}^{\rm SrTiO} - nE_{bulk} )
\end{eqnarray}
where $E_{slab}^{i}$ the total-energy of the symmetric slab with the $i$ termination, $n$ the total number of bulk formula units in the two slabs, $S$ the surface area and E$_{bulk}$ the bulk energy per formula unit in the cubic structure.

Since the (110)-O termination is self-complementary and shows an AFD distortion, its cleavage energy $E_{cl}^{(O + O)}$ is calculated by using a unit factor instead of one half and a different reference energy E$_{bulk}^{AFD}$, which reads:
\begin{equation}
E_{cl}^{\rm O+O}=\frac{1}{S}(E_{slab}^{\rm O} - nE_{bulk}^{AFD})
\end{equation}
As previously pointed out in Section II, a different choice for the reference energy would result in a ill-defined cleavage energy. 
This is exemplified in a recent calculation, in which an apparently diverging cleavage energy was obtained for the (110)-O termination~\cite{HeifetsSS} by using the cubic, undistorted reference energy instead of $E_{bulk}^{AFD}$.
Our numerical results are summarized in table~\ref{Tab:Esurf}. 
The O termination has the lowest cleavage energy (2.54 J/m$^2$) whereas the O$_2$ and SrTiO the highest one (6.52 J/m$^2$).
We also note that the three non-stoichiometric terminations have lower cleavage energies than the two stoichiometric ones.
Consequently, the polarity compensation that is achieved through the  modification of the surface stoichiometry seems to be more effective than that by the anomalous filling of the surface states, as far as the energetics is concerned.

\begin{table}[h]
\caption{ The cleavage energies E$_{cl}$ in J/m$^2$.}
\begin{center}
\begin{tabular}{cccc}
\hline
\hline
& O + O & TiO + Sr & SrTiO + O$_2$   \nonumber \\
\hline
E$_{cl}$ & 2.54  & 3.86     & 6.52   \nonumber \\
\hline
\hline
\end{tabular}
\end{center}
\label{Tab:Esurf}
\end{table}

\subsection{The surface grand-potential.}

In order to distinguish the contribution of each termination to the cleavage energy, we compute its surface grand potential, which implies a contact with matter reservoirs. 
Many authors have recently used this method with success, as well for binary~\cite{WangPRLal2o3,WangPRLfe2o3} as for ternary compounds~\cite{VanderbiltSSsrtio3,ArianaSS}.
We introduce the chemical potential $\mu_{\rm Ti}$,
$\mu_{\rm Sr}$ and $\mu_{\rm O}$ of the Ti, Sr and O atomic species, respectively. 
The surface Grand Potential per area unit $\Omega^i$ of the $i$ termination reads \endnote{The -TS term is neglected for typical temperatures T (S is the entropy of the system)}:
\begin{equation}
\Omega^i = \frac{1}{2S}[ E_{slab}^i - N_{\rm Ti}\mu_{\rm Ti} - N_{\rm Sr}\mu_{\rm Sr} - N_{\rm O}\mu_{\rm O}]
\label{eq:Gpot}
\end{equation}

with $N_{\rm Ti}$, $N_{\rm Sr}$ and $N_{\rm O}$ the number of Ti, Sr and O atoms in the slab and the half factor corresponds to the surface Grand potential by termination.
The chemical potential $\mu_{\rm SrTiO_3}$ of a condensed and stoichiometric phase of strontium titanate is written as a sum of three terms representing the chemical potential of each species within the crystal:
\begin{equation}
\mu_{\rm SrTiO_3}=\mu_{\rm Sr} + \mu_{\rm Ti} + 3\mu_{\rm O}
\label{eq:develop}
\end{equation}
Since the surface is in equilibrium with the bulk SrTiO$_3$, we have $\mu_{\rm SrTiO_3}=E_{bulk}$.
If we replace equation~\ref{eq:develop} in equation~\ref{eq:Gpot}, we can eliminate the $\mu_{\rm Ti}$ and $\mu_{\rm SrTiO_3}$ variables in the surface Grand Potential and obtain:
\begin{equation}
\Omega_s^i=\frac{1}{2S}[ E_{slab}^i-N_{\rm Ti}E_{bulk} - \mu_{\rm O}(N_{\rm O}-3N_{\rm Ti}) - \mu_{\rm Sr}(N_{\rm Sr}-N_{\rm Ti})]
\label{eq:Gpotnew}
\end{equation}

Relying upon equation \ref{eq:Gpotnew}, one can deduce, for each termination, the range of the accessible values of $\Omega^i_s$ if the minimum and maximum values of the O and Sr chemical potentials are known (see appendix).

If we introduce the variation of the chemical potentials with respect to those computed for the reference phases ($\Delta \mu _{\rm O} = \mu _{\rm O} - \frac{E^{mol}_{\rm O_2}}{2}$ and $\Delta \mu _{\rm Sr} = \mu _{\rm Sr} -E^{bulk}_{\rm Sr}$, respectively) in equation~\ref{eq:Gpotnew} we obtain:

\begin{eqnarray}
\Omega_s^i & = & \phi_i - \frac{1}{2S}[\Delta \mu_{\rm O}(N_{\rm O}-3N_{\rm Ti})
-\Delta \mu_{\rm Sr}(N_{\rm Sr}-N_{\rm Ti})] \nonumber \\
\mbox{with     } \phi_i & = &  \frac{1}{2S} [ E_{slab}^i-N_{\rm Ti}E_{bulk}- \nonumber \\
& &\frac{E_{\rm O_2}^{mol}}{2}(N_{\rm O}-3N_{\rm Ti})
-E_{\rm Sr}^{bulk}(N_{\rm Sr}-N_{\rm Ti}) ] 
\label{eq:Gpotter}
\end{eqnarray}
$\phi _i$ measures the stability of the surface with respect to bulk SrTiO$_3$, gaseous molecular oxygen and metallic Sr. In Table~\ref{Tab:gamma} our {\it ab initio} results for $\phi _i$, are listed.

\begin{table}[h]
\caption{The surface energy $\phi_i$ as defined in Eq.~\ref{eq:Gpotter} 
is given for each termination $i$ in J/m$^2$.}
\begin{center}
\begin{tabular}{cccccc}
\hline
\hline
$i$ & O & TiO & Sr  & SrTiO & O$_2$   \nonumber \\
         \hline
$\phi_i$ & 1.27  & 6.88&-3.01&  4.78 & 1.73 \nonumber \\
\hline
\hline
\end{tabular}
\end{center}
\label{Tab:gamma}
\end{table}

For two complementary terminations (TiO-Sr and SrTiO-O$_2$) the sum of their surface Grand potential is independent of the chemical potential and corresponds to their cleavage energy (see Table~\ref{Tab:Esurf}).

The derivation of the upper and lower bounds of $\Delta \mu_{\rm O}$ and $\Delta \mu_{\rm Sr}$ is detailed in appendix A.
Within the allowed region, we show in figure~\ref{Fig:Stab110} (left panel) the $(1\times 1)$ (110) terminations having the lowest surface grand potentials, which provides the stability diagram of SrTiO$_3$(110) $(1\times 1)$ surfaces in an O and Sr external environment.
\begin{widetext}
\begin{center}
\begin{figure}[h!]
\mbox{\includegraphics[height=18cm,width=6.5cm,angle=-90]{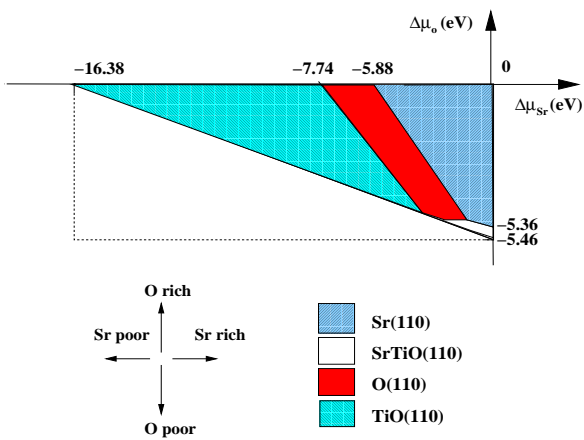}}  
\caption{{Stability diagram of the of the (1$\times$1)
SrTiO$_3$ (110) surface. The actual most stable termination is
represented in the left panel as a function of the excess O and Sr chemical 
potentials $\Delta \mu _o$ (vertical) and $\Delta \mu _{Sr}$ (horizontal). In the 
right panel, the surface Grand potentials are represented as a function of $\Delta \mu_O$ 
(for the particular value of the Sr chemical potential $\Delta \mu_{Sr}=0$ eV).}}
\label{Fig:Stab110}
\end{figure}
\end{center}
\end{widetext}

First of all, according to Fig.\ref{Fig:Stab110} (left panel), our calculations predict that only four out of the five possible terminations may be obtained. Indeed, the (110)-O$_2$ termination cannot be stabilized, even in very O-rich chemical environments. 
The Sr termination is the most stable one in O- and Sr-rich environments, as its complementary TiO face is in O- and Sr-poor conditions. 
The O termination shows a stability domain in moderate O- and Sr-environment.
Finally, the stoichiometric and open-shell (110)-SrTiO termination happens to be stable in a small domain corresponding to O-poor and Sr-rich conditions. In order to dicuss the existence of this small domain (at least within the theory) with respect to the precision of our calculations, we point out that (see the right panel of Fig. 11):
\begin{itemize}
\item At the O-poor and simultaneously Sr-rich zone boundary (i.e.: $\Delta \mu_O$=-5.46 eV and $\Delta \mu_Sr$= 0 eV) the difference between Sr and SrTiO surface Grand Potentials is equal to 0.13 J/m$^2$, which is higher than the estimated precision of our calculations, which is of the order of 0.01 J/m$^2$.
\item Since the area of the ($\Delta \mu_O$, $\Delta \mu_Sr$) domain is slightly underestimated by our first principle calculations (see appendix), the consideration of the experimental boundaries would enlarge the stability domain of the SrTiO termination.
\end{itemize}

\section{Discussion}

Among the questions that we raise in the introduction, in the following we discuss three issues that are especially noteworthy.

\subsection{A stable polar stoichiometric termination: (110)-SrTiO.}

We have shown in the previous section that the $(1\times 1)$ stoichiometric (110)-SrTiO termination can be stabilized in a O-poor and Sr-rich environment.
Even if its stability domain comes out to be rather small, this is the second case (to our knowledge) together with the Zn and O terminations of ZnO(0001)~\cite{Wander} of a stable stoichiometric polar oxide surface with an open-shell electronic structure. 
However, a recent STM investigation \cite{Diebold02} questioned the proposed stoichiometric morphologies for ZnO(0001) and provides different, non stoichiometric, structural models.

The a polarity compensation through anomalous filling of surface states is expected on the basis of theoretical arguments, but it is actually not often encountered~\cite{Noguerajpcm}.
Indeed, the cleavage energies of stoichiometric polar terminations in rock-salt structures, such as the $(1\times 1)$ unreconstructed (111) faces of MgO, are generally much higher than those of reconstructed, non-stoichiometric polar terminations. 
As a consequence, the surface grand potentials of these polar stoichiometric terminations result higher than those of the non stoichiometric ones even when the chemical potential dependent terms represent a negative contribution. 
In this respect, the peculiar behavior of SrTiO$_3$ may be due to the presence of Ti-O covalent bonds, in conjunction with a not too large fundamental gap. 
Indeed, the energy increase that is due to the anomalous filling of surface states needed for polarity compensation may be effectively lessen by atomic relaxation and electronic screening effects.
The latter ones show up through a non negligible charge transfer affecting some surface layers as supported by the analysis of the topological Bader's charge of the SrTiO termination (Table~\ref{Tab:Bader_srtio}): all Ti and O charges belonging to the three topmost layers are modified. 
Such a screening mechanism is less costly than a drastic charge reduction on only one or two surface atoms, which is essentially the case for MgO(111) $(1\times 1)$~\cite{Finocchi-np}.

\subsection{Comparison with the non polar SrTiO$_3$(100) terminations.}

\begin{figure}[h]
\centering
\mbox{\includegraphics[height=9cm,width=7cm,angle=-90]{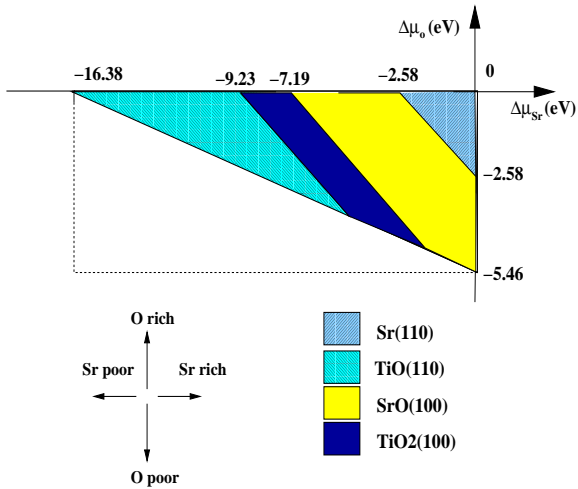}}
\caption{{Phase diagram: comparison between the different terminations
of the (1$\times$1) SrTiO$_3$ (110) SrTiO, O$_2$, TiO, Sr
and O terminations and SrTiO$_3$ (100) TiO$_2$ and SrO terminations
in an Oxygen and Strontium external environment.}}
\label{Fig:Comp110_100}
\end{figure}

The cleavage (100) orientation of SrTiO$_3$ is non-polar and displays two different stoichiometric terminations: the (100)-TiO$_2$ and the (100)-SrO. 
They are usually expected to be more stable than any polar face such as the (110) terminations.
Indeed, within the same theoretical and computational framework, we calculated the cleavage energy of the $(1\times 1)$ (100)-TiO$_2$ and -SrO surfaces, which is not very much lower than those of the non-stoichiometric (110) terminations. 
Therefore, it is worth comparing the thermodynamic stability of the simulated $(1\times 1)$ (110) terminations with the two $(1\times 1)$ (100)-TiO$_2$ and (100)-SrO faces.
In figure~\ref{Fig:Comp110_100} the domains of stability of the two (100) terminations and the five (110) terminations previously reported are gathered.
Even if in the most common conditions, corresponding to moderate Sr and O chemical potentials, the (100) faces are favored, two (110) distinct $(1\times 1)$ terminations are predicted to be stable -- the (110)-TiO in Sr-poor environments and the (110)-Sr in O-rich and Sr-rich conditions -- the first of which shows a quite wide domain of thermodynamic stability.
From the theoretical point of view, the (110)-TiO and the (110)-Sr terminations can be thought as exposing strongly relaxed \{100\} nano-facets (see Fig.\ref{Fig:Cut_srtio_tio2}). It is therefore not completely surprising that their surface energies can become comparable to those of flat \{100\} terraces, if the formation energy of corners and edges is not too high.
This is consistent with the experimental evidence, which shows that \{110\} orientations can be quite easily obtained for SrTiO$_3$.

\subsection{Comparison with experimental results.}

Using the first-principles stability diagrams displayed above, we can now discuss and propose a tentative explanation for the experimental measurements.
As anticipated in the introduction, the SrTiO$_3$(110) surface was produced and characterized by several groups, showing a great sensitivity to the preparation conditions.
In particular, if the SrTiO$_3$(110) is slightly annealed in ultrahigh vacuum (UHV), at 800${\rm ^oC}$ ~\cite{Bando2} or at 960${\rm ^oC}$ during 2 hours~\cite{Zegen}, the surface exhibits a (1$\times$1) LEED pattern. 
In contrast, a $(n\times m)$ periodicity is observed when the SrTiO$_3$(110) surface is annealed at temperatures $T \gtrsim 1000 {\rm ^oC}$. 
Therefore, one may guess that $(1 \times 1)$ SrTiO$_3$ (110) surfaces represent local minima of the possibly complicated free-energy landscape in a quite wide temperature range. 
Preliminary calculations on larger cell reconstructions \cite{Bottin-unp} show that some $(1\times 1)$ phases may remain stable with respect to other reconstructed surfaces. 
In the following, we mainly discuss the experiments performed on these $(1\times 1)$ phases.

As far as the composition of the surface is concerned, the authors of Ref.~\onlinecite{Zegen} get evidence through Auger measurements of an increasing surface concentration of Sr at the high-temperature annealed surface.
Therefore, they conclude for a lack of Strontium for the mildly annealed (1$\times$1) surface, and propose a \{100\} micro-faceted (110)-TiO$_2$ termination as the corresponding atomic-scale structural model.
On the other hand, the STM study of Ref.~\cite{Bando2} shows on the (1$\times$1) terminations the presence of rather flat regions where tunnelling spectra have metallic character. 
These findings are consistent with typical Ti$^{3+}$ and Ti$^{2+}$ features in the XPS spectra, as well as a metallic surface state with a Ti 3d character as seen in UPS.
On the basis of all those observations, and at odds with Ref.~\onlinecite{Zegen}, they propose a SrTiO termination for the unreconstructed SrTiO$_3$(110)  surface, which can account for its flatness and metallic character dominated by Ti-like occupied states.

Relying on our calculations, we argue that these two apparently contradictory models are not necessarily incompatible. 
On one hand, the UHV Auger measurements may have been carried out in Sr-poor environments, whereas the UHV experimental conditions that have been used in Ref.~\onlinecite{Bando2} imply an O-poor environment. 
Keeping in mind the great sensitivity of the SrTiO$_3$(110) surface to the actual thermodynamic conditions, it is not surprising that its atomic-scale structure may show big variations.
According to our calculations (see Fig.\ref{Fig:Stab110}), the (110)-TiO termination, which can be interpreted as a \{100\}-TiO$_2$ micro-faceted surface in agreement with the model proposed by Brunen and Zegenhagen~\cite{Zegen}, can be obtained in a wide domain corresponding to slightly Sr-poor and O-poor environments.
If we assume that the temperature annealing is done in O-poor and, at the same time, rather Sr-rich conditions (whether one starts from the very beginning with a stoichiometric SrTiO termination without desorbing Sr atoms or the latter ones migrate from the bulk to surface domains of TiO composition remains an open question) a local (110)-SrTiO termination may be obtained as proposed by Bando and coworkers~\cite{Bando2}.
Indeed, the metallic character of the (110)-SrTiO termination that is due to a surface state mainly of Ti character agrees with the measured UPS spectra.
Moreover, as we have previously pointed out for this termination, the anomalous filling of a Ti surface state that is needed for polarity compensation, should be associated to special Ti$^{n+}$ features in XPS, with $n<4$.

\section{Conclusion}

The SrTiO$_3$(110)(1$\times$1) surface was studied in the framework of {\it ab initio} calculations for the first time. 
The number of surface terminations in this ternary compound is larger than in binary compounds, which permits a remarkable variety for mechanisms of polarity compensation.
An anomalous filling of the surfaces states takes place at the SrTiO termination with an open-shell electronic structure and at the O$_2$ termination with in-gap states. 
If non-stoichiometric O, TiO and Sr terminations are considered, we obtain for each of them an insulating bulk-like electronic structure. \\

By calculating the surface grand potentials, we obtain four distinct $(1\times1)$ stable (110) terminations as a function of the Sr and O chemical potentials. 
A quite large domain is found for the three non-stoichiometric terminations, especially for the TiO one.
This competition between different terminations is also confirmed by the presence of a small domain of stability for the SrTiO termination in simultaneously Sr-rich and O-poor environments.  \\

In conclusion, due to the scarcity of experimental investigations and the observed complexity of the physical and chemical behaviors of the SrTiO$_3$(110) surface, the detailed determination of its atomic and electronic structure still remains an open question. 
However, our first-principles calculations suggest that even for the unreconstructed $(1\times 1)$ surfaces many distinct terminations are likely to appear, according to the precise experimental conditions. 
In particular, the great dispersion of the experimental results should be connected with such a sensitivity. 
In the light of our simulations, the available structural models that have been proposed to interpret the measurements seem to be reasonable within a restricted thermodynamic domain. 
We point out that coupling spectroscopic measurements and structure-sensitive techniques such as near-field microscopies and grazing x-ray diffraction, in very carefully controlled chemical environments may be crucial to get a comprehensive insight into SrTiO$_3$ polar surfaces.   


\begin{acknowledgments}
We thank P. Casek for helping us in performing the Bader's topological charge analysis. 
We also ackowledge financial support from the GDR CNRS "Interface et Surface Sensible \`a la Structure"
and from the Laboratoire de Physique des Solides in Orsay, France. 
The calculations were done on the IBM RS/6000 SP Power3 computer at IDRIS, under project No. 24089. 
\end{acknowledgments}

\appendix
\section{Boundary limits for the chemical potentials}

In this appendix we derive the range of accessible values for $\mu_{\rm Sr}$ and $\mu_O$ in the stability diagram of the SrTiO$_3$(110) orientation. 
The Oxygen, Titanium and Strontium atoms are assumed to form no condensate on the surface. 
Consequently, the chemical potential of each species must be lower than the energy of an atom in the stable phase of the considered species:

\begin{eqnarray}
\label{eq:ineq1}
\Delta\mu_O & = & \mu_O -\frac{E_{\rm O_2}^{mol}}{2} < 0 \\
\label{eq:ineq2}
\Delta\mu_{\rm Sr} & = & \mu_{\rm Sr} - E_{\rm Sr}^{bulk} < 0 \\
\label{eq:ineq3}
\Delta\mu_{\rm Ti} & = & \mu_{\rm Ti} - E_{\rm Ti}^{bulk} < 0
\end{eqnarray}

\begin{figure}[h]
\centering
\mbox{\includegraphics[height=9cm,width=7cm,angle=-90]{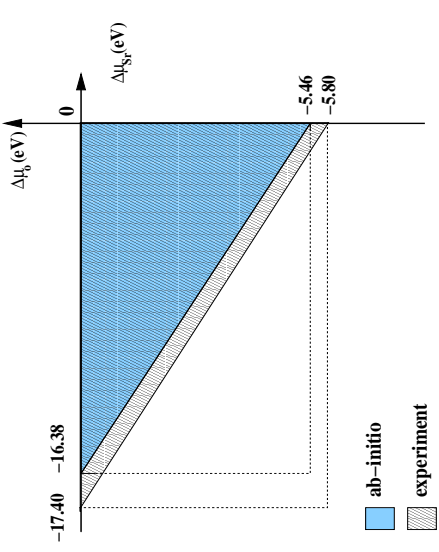}}
\caption{{Stability of the Strontium Titanate phase
in a ($\Delta\mu_{\rm Sr}$;$\Delta\mu_{\rm O}$) plane.}}
\label{Fig:stability}
\end{figure}

We have introduced in these inequations the relative values $\Delta\mu_O$, $\Delta\mu_{\rm Sr}$ and $\Delta\mu_{\rm Ti}$ of the different chemical potentials with respect to $E_{\rm Sr}^{bulk}$, $E_{\rm Ti}^{bulk}$ and $\frac{E_{\rm O_2}^{mol}}{2}$, which are the energies of a Ti atom in the hcp bulk metal, of the Sr atom in the bulk cubic structure and of the O atom in the O$_2$ molecule in the gas phase, respectively.
The two first inequations~\ref{eq:ineq1} and ~\ref{eq:ineq2} define the upper boundaries of the Oxygen and Strontium chemical potentials.
By combining Eq.~\ref{eq:ineq3} and Eq.~\ref{eq:develop} in the main text, we obtain the following lower boundaries:
\begin{eqnarray}
\Delta\mu_{\rm Sr}+3\Delta\mu_O & > & -E_{\rm SrTiO_3}^f \mbox{  with} \\
-E_{\rm SrTiO_3}^f & = & E_{\rm SrTiO_3}^{bulk}-E_{\rm Ti}^{bulk}-E_{\rm Sr}^{bulk}- \frac{3}{2}E_{\rm O_2}^{mol} \nonumber
\label{eq:thirdineq}
\end{eqnarray}
$E_{\rm SrTiO_3}^f$ is the formation energy of SrTiO$_3$ with respect to the Ti and Sr atoms in their bulk phases, and the O atom in the gas phase wich is positive defined. 
In order to compute easily this quantity and compare it with the experimental values, we rewrite $E_{\rm SrTiO_3}^f$ in the following way:
\begin{equation}
E_{\rm SrTiO_3}^f= E_{coh}-E_{\rm Sr}^f-E_{\rm Ti}^f-\frac{3}{2}E_{\rm O_2}^f
\end{equation}
with all these quantities defined in Sec.~\ref{Computational}.
Our computed value is $E_{\rm SrTiO_3}^{f}=16.38$ eV to be compared to the value $E_{\rm SrTiO_3}^{f}=17.40$ eV that can be deduced by the experimental data. \\
In conclusion we show the stability of the Strontium Titanate diagram in a ($\Delta\mu_{\rm Sr}$;$\Delta\mu_{\rm O}$) plane in Fig.~\ref{Fig:stability}.
As  one can see, the {\it ab initio} calculations only slightly underestimate the size of the stability region.


\newpage 
\bibliographystyle{apsrev}
\bibliography{version6}

\end{document}